\documentclass[doublecol]{epl2}

\title{Fractoluminescence characterization of the energy dissipated during fast fracture of glass}
\shorttitle{Fractoluminescence characterization of the energy dissipated during fast fracture}

\author{G. Pallares\inst{1,2} \and C. L. Rountree\inst{1} \and L. Douillard\inst{1} \and F. Charra\inst{1} \and E.
Bouchaud\inst{3,4}}
\shortauthor{G. Pallares \etal}

\institute{
  \inst{1} CEA Saclay, IRAMIS, SPCSI - Building 462, F-91191 Gif-sur-Yvette Cedex, France, EU\\
  \inst{2} LPMCN, Universit\'e Lyon 1, CNRS - 43 Boulevard du 11 Novembre 1918, F-69622 Villeurbanne, France, EU\\
  \inst{3} CEA Saclay, IRAMIS, SPEC - Building 772, Orme des Merisiers, F-91191 Gif-sur-Yvette Cedex, France, EU\\
  \inst{4} ESPCI-Paris Tech, PSL$^{\star }$, UMR Gulliver, EC2M, 10 rue Vauquelin, F-75231 Paris Cedex 05, France, EU\\}


\abstract{Fractoluminescence experiments are performed on two kinds of
silicate glasses. All the light spectra collected during dynamic
fracture reveal a black body radiator behaviour, which is
interpreted as a crack velocity-dependent temperature rise close
to the crack tip. Crack velocities are estimated to be of the
order of 1300~m.s$^{-1}$ and fracture process zones are shown to extend
over a few nanometers.}

\begin{document}

\maketitle

Although glass is considered as the archetype of brittle
elastic materials, it was shown in various situations that, during
fracture, a large part of the stored elastic energy is dissipated in
permanent deformation of the material. It was argued to be mostly
dissipated in the formation of nanocracks in ultraslow stress
corrosion cracking conditions~\cite{Celarie2003,Lechenault2011}. In
vacuum, Molecular Dynamics (MD) simulations~\cite{Rountree2007} of
dynamic fracture predict that energy dissipation results both from
plastic deformation and from bond breaking
 ahead of the crack tip, the two phenomena acting at the scale of a few nanometers.
The latter predictions are in agreement with experiments
due to S. Wiederhorn \cite{Wiederhorn1969}, who performed
accurate measurements of the energies of fracture  in an
inert environment for different glass compositions, and
showed that they are approximately ten times higher than
the typical surface energy values estimated by Griffith
\cite{Griffith1920}. However, in order to understand fully
the nature of the dissipation processes, it is highly
desirable to measure the size of the Process Zone (PZ)
where they take place for a given crack velocity. Direct
measurements being hardly imaginable because of the small
dimensions involved, and because crack velocities are
larger than a few hundreds of meters per second, one has to
resort to indirect characterizations. The scope of the
present work is to show how fractoluminescence can be used
for that purpose.

During rapid fracture, the emission of neutral particules, ions,
electrons and photons has indeed been observed in a wide variety of
materials \cite{Walton1977,Dickinson1981,Bahat1985}. Fractoemission has been
used to probe both fracture mechanisms and fracture surface chemistry
~\cite{Pantano1985}. Since an early paper by
Wick et al. \cite{WicK1937}, a growing interest has fostered in
oxide glass fractoluminescence, and the first wavelengths spectra
were obtained in the eighties \cite{Zink1982}.

These spectra usually exhibit both an energy continuum and peaks
corresponding to discrete energy values. Chapman and
Walton~\cite{Chapman1983} performed dynamic fracture experiments on
different glasses, and showed that the continuum corresponds to a
black body radiator spectrum.  Using a model developped by Weichert
and Schonert~\cite{Weichert1978}, they compared the numerical
solutions obtained by these authors to their own observations in
order to evaluate the size of the heated propagating zone ahead of
the crack tip, and found  it to be of the order of a few nanometers.
More recently, Gonzalez and coworkers~\cite{Gonzalez1990} showed
that no photons were detected for crack velocities smaller than
$10^{-2}$~m.s$^{-1}$ in soda-lime silica.

The central point of this letter is to provide quantitative evidence
for the above scenario by interpreting the black body emission
spectrum observed during dynamic fracture within the framework of
Rice and Levy's~\cite{Rice1969} and Freund's~\cite{Freund1990}
models. We estimate both the crack velocity and the size of the PZ,
where dissipative processes occur and the temperature rises.

\section{Experimental Setup}
\begin{figure}
\centering
\includegraphics[width=0.42 \textwidth]{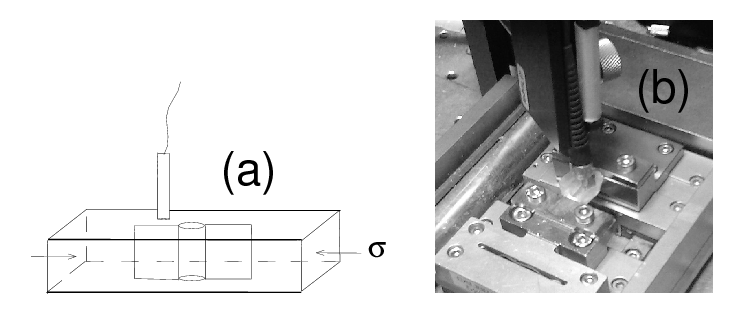}
\caption{Experimental setup: (a) Sketch of the DCDC geometry: two
symmetric cracks propagate from the central hole, and the photon
sensor is placed on the path of one of them; (b) Photograph of the
setup.} \label{fig1}
\end{figure}

\begin{figure}[!h]
\begin{center}
\includegraphics[width=.40 \textwidth] {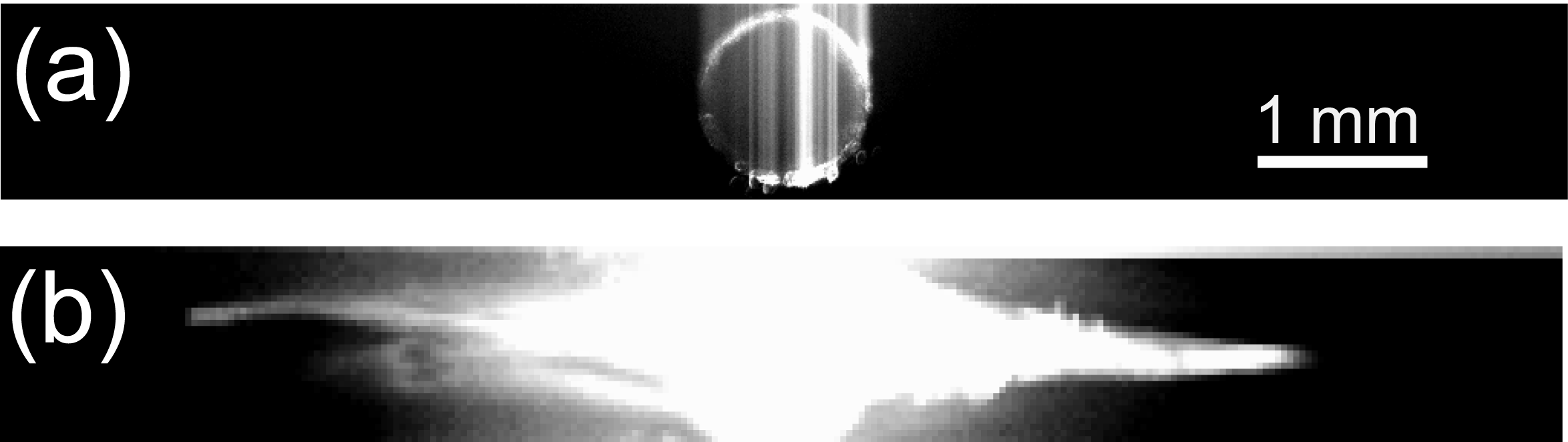}
\caption{Imaging-mode measurement during dynamic fracture on float
glass sample. The top and the bottom images  are acquired
respectively before and during fracture. Two cracks emerge
symmetrically from the hole, starting by being quite straight - as
observed in \,(b) - before the specimen shatters.} \label{fig2}
\end{center}
\end{figure}

Double cleavage drilled compression
(DCDC) specimens (Fig.~\ref{fig1}) with implemented precrack obtained by stress corrosion~\cite{Prades2005}
are used to perform experiments. Cracks are propagated at a constant temperature $T_0 = (300
\pm 2)$~K under ambient pressure. Samples made of pure silica (Corning
7980, USA) and float glass (Saint Gobain, France) are
parallelepipedic ($5 \times 5 \times 25$~mm$^3$, with 10~$\mu$m
tolerance). A hole of radius $a=(500 \pm 10)$ $\mu$m is drilled in
their center to trigger the formation of two symmetric mode I
tensile cracks sketched in Fig.\ \ref{fig1}\,(a) when the sample is
submitted to uniaxial compression at imposed displacement speed~\cite{He1995,Fett2005,Pallares2009}.
To avoid stress corrosion, the specimen is placed in a bath of dodecane
oil preventing the water molecules penetration at the crack tip.
This results in dynamic fracture as suggested by
Wiederhorn~\cite{Wiederhorn1974}. Potential light intrusion has been
identified and neutralized.
In order to check that cracks actually propagate within the midplane
of the sample, we first monitor experiments by imaging the emitted
photons using a large-aperture lens (numerical aperture: 0.4, focal
length: 12mm) onto the 1024x128 pixel array (26.6~mmx3.3~mm) of a
cooled (193~K) CCD camera (Andor, DU-401 BR-DD) with a 1:2.75
magnification factor. In accordance with standard Charge-Coupled Device (CCD) operation, the radiation is continuously integrated and stored as photo-generated charges in the pixels of the CCD, until a read-out operation of the accumulated charges takes place. In order to reduce the level of dark noise, we programmed a read-out every 5 seconds. This period has been optimized in order that the accumulated dark noise ($\sim $~4 photo-electrons per pixel) remains smaller than the signal emitted by a single crack event, whereas the probability that the emission by the short-lived crack event occurs during a read-out ($\sim $~50 ms) is negligible.
 The detection spectral range is nominally 400~nm -- 1150~nm.
Fig.\ \ref{fig2}\,(a) and \,(b) shows two $1.2
\times 9.3$~mm$^2$ images which are
acquired respectively before and during dynamic fracture.

In order to measure the photon spectra, a 300-$\mu $m-diameter fiber
bundle is used to collect the emission locally on the propagation
path on the sample at $d=1.0\pm0.2$~mm from the hole as shown in
Fig.\ \ref{fig1}\,(a). The output of the fiber bundle forms the
entrance slit of a spectral disperser (Andor, Shamrock 163i) coupled
to the above-mentioned cooled CCD detector. The spectral resolution
is 4~nm in wavelength.

\section{Experimental results}
\begin{figure}[!h]
\begin{center}
\includegraphics[width=.47 \textwidth] {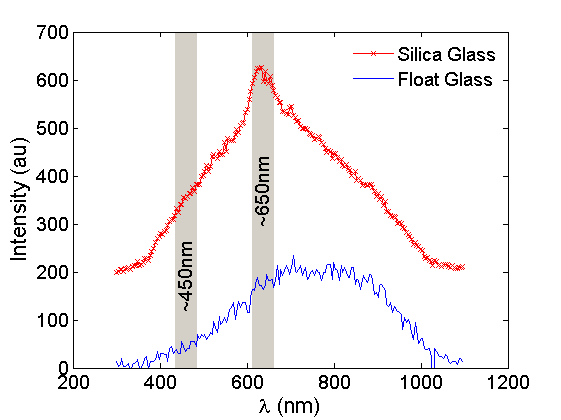}
\caption{Experimental spectra obtained on pure silica (red x) and on
float glass (blue line) during dynamic fracture. The 450~nm ($\sim$ 2.7-eV) and
650~nm ($\sim$ 1.9-eV) bands are enlighted in gray.} \label{fig3}
\end{center}
\end{figure}

Fig.~\ref{fig3} shows typical spectra
obtained for silica and float glass. The signal intensity is
always observed to be smaller for float glass, for which the maximum
consistently takes place at lower energy values. On the pure silica
spectra, we note, on top of a continuous spectrum, the presence of a
characteristic peaks corresponding to nonbridging oxygen hole center
relaxation at 650~nm ($\sim$ 1.9-eV)~\cite{Tohmon1989}.

\begin{figure}[!h]
\centering
\includegraphics[width=0.47 \textwidth]{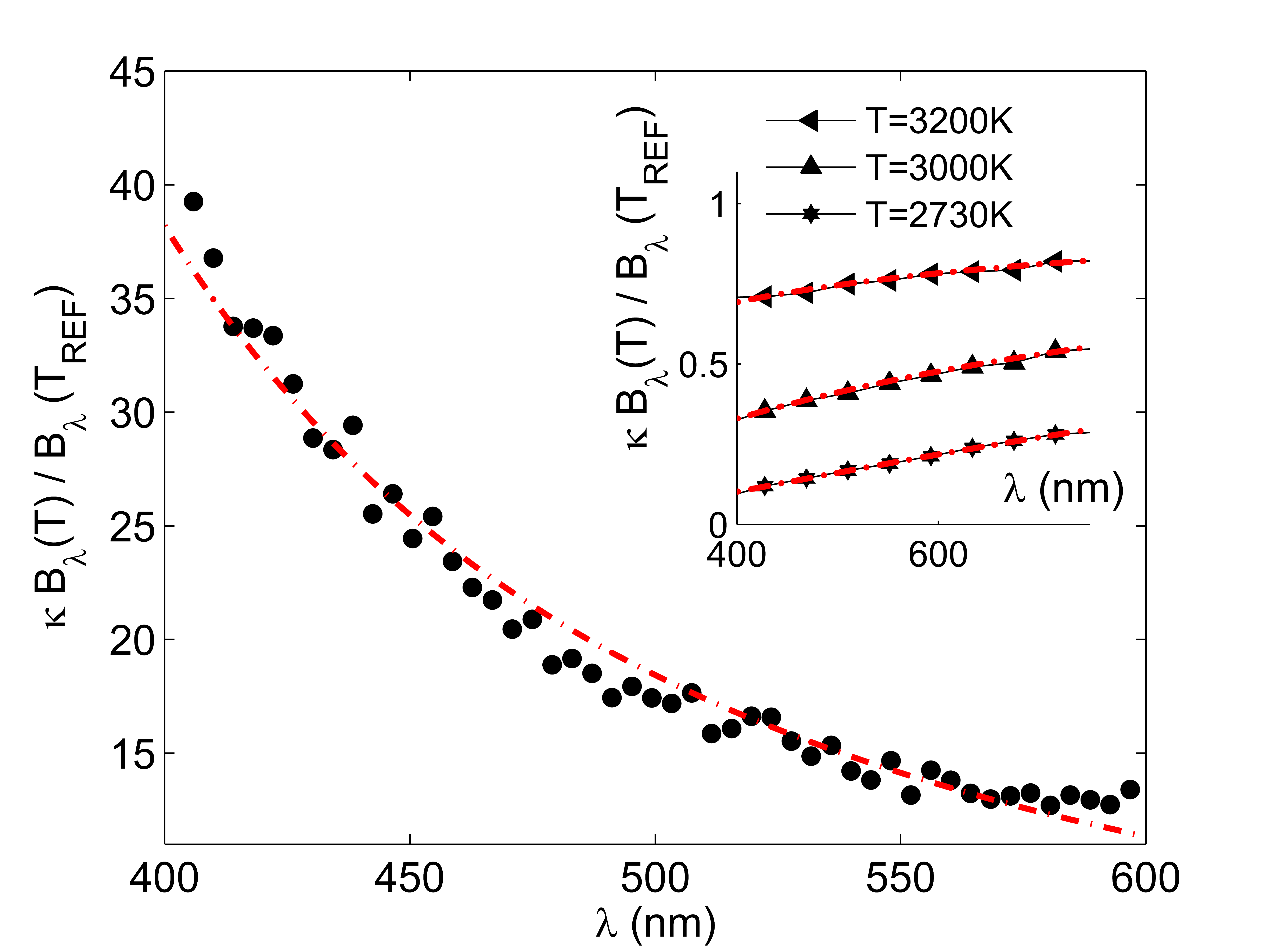}
\caption{Experimental renormalized spectrum ratio $\kappa B_{\lambda
}(T)/B_{\lambda }(T_{REF})$ obtained on pure silica glass during
dynamic fracture. The crack tip temperature is found close to
$T\simeq5000$~K -- Inset: Example of three known temperature
spectra ratio obtained by the thermal source (Schott KL 1500
LCD).} \label{fig4}
\end{figure}

In order to check that the continuous part of each observed spectrum
is black body type, and to evaluate the corresponding temperature,
each curve is normalized by the spectrum of a thermal source with
known temperature (Schott KL 1500 LCD). This procedure permits to
get rid of the overall spectral sensitivity function of the
detection system. According to \cite{Soffer1999}, the Planck
distribution function in terms of photons number per second and per
unit wavelength $\lambda$ can be expressed as $B_{\lambda}(T)
\propto \frac{\lambda^{-4}}{e^{\beta/\lambda T}-1}$ where
$\beta=hv_{l}/k_B\simeq 0.01439$~m.K ($h, v_{l}$ and $k_B$ being
respectively the Planck constant, the velocity of light and the
Boltzman constant).The measured renormalized spectra were fitted in
the wavelength domain 400--700~nm by the ratio $\kappa B_{\lambda
}(T)/B_{\lambda }(T_{REF})$, where $\kappa$ is a constant accounting
for the difference in duration and spatial extension of the sources.
This temperature evaluation method was first tested on five known
temperature spectra obtained by the thermal source ranging between
2650 and 3200~K. Note that our detector fails to detect IR thermal photons, because of the fast thermal diffusion following strongly localized heating, and the subsequent temperature decay: the IR emission persists longer than the UV-Vis one.  Thus the fit concerns only the range 400-740 nm in this case. The relative error was found to be less than 1\%
(cf the inset of Fig.~\ref{fig4}). For our glass samples, the fitting range was further reduced to 400-600nm because of the existence of a peak at 650nm. Temperature rises at crack tips,
both in silica and float glass samples, are given in Table~\ref{tab:results}.

\section{Local heating induced by non linear deformations in dynamic
fracture}
The temperature field of a crack tip running at a constant
velocity $v_c$ was determined by Rice and Levy~\cite{Rice1969}. They
showed that the temperature rise within the PZ around the crack tip
is equal to:
\begin{equation}\label{eq:RiceLevy}
\Delta T = \frac{\sqrt{\pi}}{2} \frac{(1-\nu^2)}{E}
\frac{K_I^2}{\sqrt{\rho ck}}\sqrt{\frac{v_{c}}{R_c}}
\end{equation}
\noindent with $K_I$ the stress intensity factor (SIF). $\nu $ and
$E$ are respectively Poisson's ratio and Young's modulus, while
$\rho $, $k$ and $c$ are respectively the density, the thermal
conductivity and the specific heat capacity of the material (see
Table~\ref{tab:ParamPhotons}).
\begin{largetable} 
\caption{$\Delta T$ is the temperature elevation in the PZ and $v_c$
is the measured crack velocity corresponding. $K_I$ and $R_c$ are
respectively the stress intensity factor and the size of the PZ.
$\ell$ is the extent of the region where the temperature decreases
from very high within the PZ to its room value far from it.}
\label{tab:results}
\begin{tabular}{c   c  c  c  c  c  c}
  \hline
       & $\Delta T$&$v_c$         &$K_I$          &$R_c$&$\ell$\\
Sample & (K)       &(m.s$^{-1}$)  &(MPa.m$^{1/2}$)&(nm) &(nm)\\
    \hline
Silica 1  & 4680$\pm$130  & 1230$\pm$94 & 0.99$\pm$0.03 & 3.9$\pm$0.2 & 0.68$\pm$0.05  \\
Silica 2  & 5480$\pm$450  & 1480$\pm$350 & 1.06$\pm$0.09 & 4.4$\pm$0.7 & 0.56$\pm$0.13  \\
Silica 3  & 4980$\pm$210  & 1330$\pm$160 & 1.02$\pm$0.04 & 4.1$\pm$0.3 & 0.63$\pm$0.07  \\
Float 1   & 5180$\pm$840  & 1550$\pm$730 & 1.02$\pm$0.16 & 4.1$\pm$1.3 & 0.27$\pm$0.13  \\
Float 2   & 4100$\pm$530  & 1170$\pm$410 & 0.93$\pm$0.12 & 3.4$\pm$0.9 & 0.36$\pm$0.12  \\
 \hline
\end{tabular}
\end{largetable}

In not too fast fracture, i.e. before complex processes occur (such
as branching~\cite{Fineberg1999} or extended damage
formation~\cite{Washabaugh1995}), the crack velocity $v_c$ can be
related to $K_I$ through Freund's elastic continuum mechanics
prediction~\cite{Freund1990}:
\begin{equation}\label{eq:Freund}
\frac{v_{c}}{v_R} = 1-\left(\frac{K_{Ic}}{K_I}\right)^2
\end{equation}
\noindent where $v_R= 3350$~m.s$^{-1}$ is the Rayleigh wave speed in
glass. Indeed,  Sharon and Fineberg \cite{Sharon1999} validated this
prediction in glass, for velocities up to $v_{c}=0.42v_R\sim
1400$~m.s$^{-1}$. For smaller velocities, invoking
Eq.~(\ref{eq:RiceLevy}) and the Dugdale-Barrenblat expression for
the PZ size:
\begin{equation}\label{eq:Dugdale}
R_c = {\pi \over 8} \left({K_I\over \sigma _y}\right)^2
\end{equation}
\noindent where $\sigma_y$ is the yield stress.
Several techniques can be used to
measure it, nevertheless the value is still debated.
The yield stress ranges approximately between 9 and 12~GPa in
silica glass under UHV conditions, depending on the loading~\cite{Rountree2002,Kalia2009,Kurkjian2003}
and references therein. For oxide glasses, Wiederhorn estimated a yield stress close to
10~GPa~\cite{Wiederhorn1969} which is in good agreement both with MD simulations~\cite{Rountree2002,Kalia2009}, where $\sigma _y$ is predicted to be close to 9GPa, and with the
results obtained by Kurkjian on glass fibers~\cite{Kurkjian2003}.
Hence, $\sigma_y$ is taken equal to 10~GPa in the following.
We can extract an expression for the crack velocity $v_{c}$:
\begin{equation}\label{eq:Velocity}
v_{c} = v_R \frac{\alpha \Delta T^2}{K^2_{Ic} + \alpha \Delta T^2}
\end{equation}
\noindent where
$\alpha= E^2 \rho c k / 2 v_R \sigma_y^2 (1-\nu^2)^2$.

Using Eq.~(\ref{eq:Velocity}), we calculate the crack velocity in
each case. The error bar $\delta v_c $ on $v_c$ can be deduced from
the error bar $\delta(\Delta T)$ on $\Delta T$: $\delta
v_c/v_c=2\delta(\Delta T)/\Delta T (1+v_c/v_R)$. This leads to the
results reported in Table~~\ref{tab:results}.
\begin{table} [h]
\caption{Values at 300~K of parameters used in
Eq.~(\ref{eq:Velocity}) to evaluate the dynamic crack velocity
$v_c$.}
\label{tab:ParamPhotons}
\begin{tabular}{c  c  c }
  \hline
 Oxide glass type &  Silica &  Float  \\
  \hline
Young modulus $E$ (GPa)  &  72.7 & 72  \\
Poisson's ratio $\nu$    &   0.16 &   0.23 \\
Density $\rho$ (kg.m$^{-3}$)  &    2201 &   2530  \\
Specific capacity $c$ (J.kg$^{-1}$.K$^{-1}$)    &  703 &   880 \\
Heat conductivity $k$ (W.m$^{-1}$.K$^{-1}$)    &  1.30 &   0.937 \\
Fracture toughness $K_{Ic}$ (MPa.m$^{1/2}$)    &  0.794 &   0.749 \\
 \hline
\end{tabular}
\end{table}

\section{Discussion}
We have confirmed  Chapman and Walton's
observations~\cite{Chapman1983}: photons emitted during dynamic
crack propagation in silicate glasses are mostly of thermal origin,
indicating a high temperature rise in a small zone around the crack
tip, where dissipative processes occur.

For pure silica, however, a peak at 650~nm corresponding to the relaxation of the nonbridging
oxygen hole center was observed~\cite{Kawaguchi1996}. On the contrary, the 450~nm signature of the
relaxation luminescence of the oxygen deficient center was not
observed in our experiments, probably because of the elevated
OH-content of our material (close to
1000~ppm)~\cite{Tohmon1989a,Kawaguchi1996}.

However, unlike Chapman and Walton~\cite{Chapman1983}, we have made
no {\it a priori} assumption on the value of the applied stress
intensity factor, the process zone size or the crack velocity. Using
Rice and Levy's model~\cite{Rice1969}, Freund's elastic
prediction~\cite{Freund1990} of the crack velocity dependence on the
fracture energy and a Dugdale-Barrenblatt expression of the plastic
zone size, we have derived a univocal relation between the
temperature rise and the crack velocity. For pure silica and for
float glass, dynamic crack velocities are found to be lying between
1172 and 1551m.s$^{-1}$ however, values exceeding the limit velocity
$\sim $1400m.s$^{-1}$ are extrapolated and may not be trusted
(Silica2 and Float1 samples in Table~\ref{tab:results}).

Moreover, since the crack propagates in the sample midplane, we
can estimate the value of the stress intensity factor $K_{Im}$ for
a crack length equal to the distance $d$ between the hole and the
detector from the measured load to fracture \cite{Pallares2009}.
In our DCDC geometry, the normalized crack length $d/a=2$ is
slightly lower than the domain where the normalized SIF equation
was adjusted, it is thus reasonable to extrapolate for estimate
purpose the stress intensity factor values: For the three silica
samples, we get $K_{Im}^{Si02}=0.98, 1.03, 1.00$~MPa.m$^{1/2}$,
and in the float glass cases $K_{Im}^{Float}=1.03,
0.92$~MPa.m$^{1/2}$. The uncertainty on $K_{Im}$ being estimated
to 0.05~MPa.m$^{1/2}$, the macroscopic values are in excellent
agreement with the instantaneous values of $K_I$ estimated
independently from the temperature elevation.

Plastic zone sizes were shown to be $\sim$~3 to 4~nm in all cases,
which is a reasonable order of magnitude~\cite{Pallares2010,Ciccotti2009}.
Note however that Molecular Dynamics simulations performed on pure silica for a crack
velocity $v_c\simeq 300$~m.s$^{-1}$  predict a slightly larger
value, close to $\sim $~10~nm, with a yield stress
of 9~GPa~\cite{Rountree2007,Kalia2009}. This can seem in
disagreement with our observations, since in dynamic fracture the
plastic zone size is supposed to increase with the applied stress
intensity factor, and hence with the crack velocity. This apparent
discrepancy can be due to a different way of estimating the PZ
size. In Rountree et al's work~\cite{Rountree2007}, the whole region
where flaws appear is taken into account.  In our case, we measure
the size of the region where the temperature rise is close to its
maximum. However, although there is actually a temperature profile
within the process zone, which could be estimated for metallic
materials~\cite{Mason1993}, we can show that for glass, the profile
is quite abrupt. As a matter of fact, what controls the extent of
the region where the temperature decreases from it elevated value
within the PZ to its room value far from it is the ratio $\ell
=k/\rho c v_c $ of the thermal diffusion coefficient $k/\rho c$ to
the crack velocity $v_c$. This length scale can be shown to be equal
to $\sim 0.6$~nm ~and $\sim 0.3$~nm~ for silica and float glass
respectively. In both cases, $\ell $ is much smaller than the PZ
size $R_c$, which means that we can consider that only the PZ is at
high temperature while the surrounding material is at room
temperature.


\acknowledgments
We are indebted to B. Pant who started these experiments
and to T. Bernard for his technical help. The authors also thank  J.-P. Bouchaud, C. Fiorini, R.K. Kalia and
K.-I. Nomura for
enlightening discussions.

\end{document}